\documentclass[%
 reprint,
 amsmath,amssymb,
 aps,
]{revtex4-1}

\usepackage{graphicx}
\usepackage{dcolumn}
\usepackage{bm}

\begin{document}

\title{A gauge-invariant, rotor Hamiltonian from dual variables of 3D $U(1)$ gauge theory}

\author{Judah F. Unmuth-Yockey}
\affiliation{Department of Physics, Syracuse University, Syracuse, NY 13244, USA}

\date{\today}

\begin{abstract}
We present a tensor formulation for free compact electrodynamics in three Euclidean dimensions and use this formulation to construct a quantum Hamiltonian in the continuous-time limit.  Gauge-invariance is maintained at every step and the resulting Hamiltonian can be written as a rotor model.  The energy eigenvalues for this Hamiltonian are computed using the tensor formulation, and compared with perturbation theory.  We find good agreement between the calculations demonstrating a smooth passage from the statistical lattice Lagrangian description to the quantum Hamiltonian description.
\end{abstract}

\maketitle

\section{Introduction}
In the last decade there has been an effort into the development and application of tensor real-space renormalization group methods for the lattice (for instance \cite{PhysRevLett.99.120601,PhysRevB.86.045139,PhysRevLett.103.160601,PhysRevD.88.056005,PhysRevLett.115.180405,doi:10.1093/ptep/ptx080} ) or TRG.  The TRG allows one to carry out genuine real-space renormalization group steps exactly as Kadanoff \cite{PhysicsPhysiqueFizika.2.263} and Wilson \cite{PhysRevB.4.3174} prescribed, and many approximation schemes have been invented within this framework.  These methods present a number of advantages over traditional sampling (Monte Carlo or MC) methods, most notably an indifference to the sign problem \cite{PhysRevD.89.016008}, and when translation invariance holds the infinite-volume limit is easily achieved.  However, it has been difficult to construct efficient TRG methods in spacetime dimensions larger than two.

An additional pleasant feature of the TRG formalism is the typical reformulation of the model of interest in terms of discrete fields, which are easier to accommodate computationally.  This approach has not only been useful in tensor formulations, but also in sampling methods \cite{GATTRINGER2018344,PhysRevD.97.034508}.  This discreteness has been found to be advantageous for making contact with quantum computation, specifically analogue quantum computing \cite{PhysRevA.90.063603,PhysRevA.96.023603,PhysRevD.92.076003}. There, one works with atomic species whose Hamiltonian descriptions are in terms of creation and annihilation operators and whose occupations are discrete \cite{Wiese:2013}.  The TRG then emerges not only as a computational tool but as a link to future computational architecture.

Here we use a particular TRG scheme, the higher-order tensor renormalization group (HOTRG) \cite{PhysRevB.86.045139}, to study the continuous-time behavior of three-dimensional compact free electrodynamics.  By reformulating this model in terms of its dual variables, we are able to rewrite the partition function as a spin-model while maintaining gauge invariance. In Ref.~\cite{kaplan:2018}, duality also plays a similar role in restricting the physical state space and enforcing Gauss's law.  We then take the continuous-time limit of this formulation as worked-out in Ref.~\cite{PhysRevD.11.395,PhysRevD.17.2637}.  The integer dual variables on the fully discrete lattice theory can be interpreted as the $z$-component angular momentum quantum numbers in the continuous-time limit, and we construct a rotor Hamiltonian for this model.

The paper is organized as follows: In Sec.~\ref{sec:dual} we introduce the model and reformulate it in terms of its dual variables.  Then using the dual variables we rewrite the partition function as a sum of local tensor contractions.  We compare with Monte Carlo to check the validity of the description.  In Sec.~\ref{sec:contime} we use the tensor formulation of the model to construct a transfer matrix and take the continuous-time limit.  In this limit we extract a quantum Hamiltonian from the transfer matrix and interpret the Hamiltonian as a rotor model.  With this Hamiltonian we compare calculations of its energy eigenvalues using the TRG with calculations done with perturbation theory and find good agreement.  Finally in Sec.~\ref{sec:conc} we give concluding remarks about the work and possible future directions.

\section{Dual variables of 3D $U(1)$ gauge theory}
\label{sec:dual}
The starting action for $U(1)$ lattice gauge theory in three Euclidean dimensions is
\begin{align}
    S &= -\beta \sum_{x \mu \nu} \Re{[U_{x,\mu} U_{x+\mu, \nu} 
    U_{x+\nu, \mu}^{\dagger} U_{x,\nu}^{\dagger}]} \\
	&= -\beta \sum_{x, \mu\nu} \cos(A_{x, \mu}
    + A_{x+\mu, \nu} - A_{x+\nu, \mu} - A_{x, \nu}) \\
    &= -\beta \sum_{x, \mu\nu} \cos(F_{x, \mu \nu}),
\end{align}
where $U_{x,\mu} = e^{i A_{x,\mu}}$ are gauge fields associated with the links of the lattice, and $\beta = 1/g^{2}$.  The partition function is
\begin{equation}
\label{eq:orig-pf}
	Z = \int \mathcal{D}[A_{x, \mu}] e^{-S},
\end{equation}
where the vector potential is periodic $A_{x,\mu} \in [0, 2\pi]$.
The Boltzmann weight can be expanded using the conjugate Fourier variables as one does in the tensor formulation, or duality transformation \cite{PhysRevD.88.056005,RevModPhys.52.453},
\begin{equation}
\label{eq:boltz-expand}
	e^{-S} = \prod_{x, \mu \nu} \sum_{n_{\mu \nu} =-\infty}^{\infty} I_{n_{\mu \nu}}(\beta) e^{i n_{\mu \nu} F_{x, \mu\nu}}.
\end{equation}
Here there is an anti-symmetric $n$ field associated with each plaquette on the lattice, and the $I_{n}(z)$ are the modified Bessel functions.  They are symmetric under $n \rightarrow -n$ for $z \geq 0$.  Each link is shared by four plaquettes in three dimensions.  The integration over the vector potential now factorizes and we find for each link,
\begin{equation}
\label{eq:guage-int}
\int \frac{d A_{x, \mu}}{2 \pi} e^{i A_{x, \mu} (n_1 + n_2 - n_3 - n_4)} = \delta_{n_1 + n_2 - n_3 - n_4, 0},
\end{equation}
where the four $n$s correspond to the four plaquettes in the co-boundary of the link.  The partition function can now be written
\begin{equation}
	Z = \sum_{\{ n \}} \left( \prod_{x, \mu\nu} I_{n_{\mu \nu}}(\beta) \right) \left( \prod_{x, \mu} \delta_{\Delta_{\nu} n_{\nu \mu},0} \right).
\end{equation}
At this point the Kronecker deltas enforcing a zero-divergence constraint can be solved identically using the curl \cite{RevModPhys.52.453},
\begin{align}
	\Delta_{\mu} n_{\mu \nu} = 0 \implies n_{\mu \nu} = \epsilon_{\mu\nu \rho}\Delta_{\rho} m.
\end{align}
The $m$s are located at the centers of the cubes of the original lattice which is necessary in order to simultaneously  satisfy all the surrounding constraints associated with the links.  Inserting this into the partition function we get,
\begin{equation}
	Z = \sum_{\{ m \}} \left( \prod_{x, \mu\nu} I_{\epsilon_{\mu\nu \rho}\Delta_{\rho} m}(\beta) \right).
\end{equation}
At this point it is convenient to switch to the dual lattice.  At the center of each cube we assign a site, and for each plaquette we assign a link connecting two dual sites.  The partition function is essentially identical,
\begin{equation}
	Z = \sum_{\{ m \}} \left( \prod_{x^{*}, \mu} I_{\Delta_{\mu} m}(\beta) \right)
\end{equation}
except the sites are the dual sites, and the product is over dual links.  We will drop the asterisk from now on and only work in the dual.  
This can be split into dual time and space links,
\begin{equation}
\label{eq:dualpf_st}
	Z = \sum_{\{ m \}} \left( \prod_{x, \tau} I_{m-m'}(\beta_{s}) \right) \left( \prod_{x, i} I_{m-m'}(\beta_{\tau}) \right).
\end{equation}
We have relaxed the notation surrounding the $m$s since the Bessel functions are symmetric in their order, and their order is the difference between $m$ values at adjacent sites.
Notice the temporal coupling is associated with the dual spatial directions and the spatial coupling is associated with the dual temporal direction.  A moment of visualization makes this clear.  
This formulation of the partition function is completely gauge-invariant, as can be seen from Eq.~\eqref{eq:boltz-expand}.  In fact, even if the sum over $n$ in Eq.~\eqref{eq:boltz-expand} is truncated, the local invariance is unaffected and one is simply left with an effective model with the same symmetries.

\subsection*{Tensor formulation of the model}

\begin{figure}[t]
    \centering
    \includegraphics[width=8.6cm]{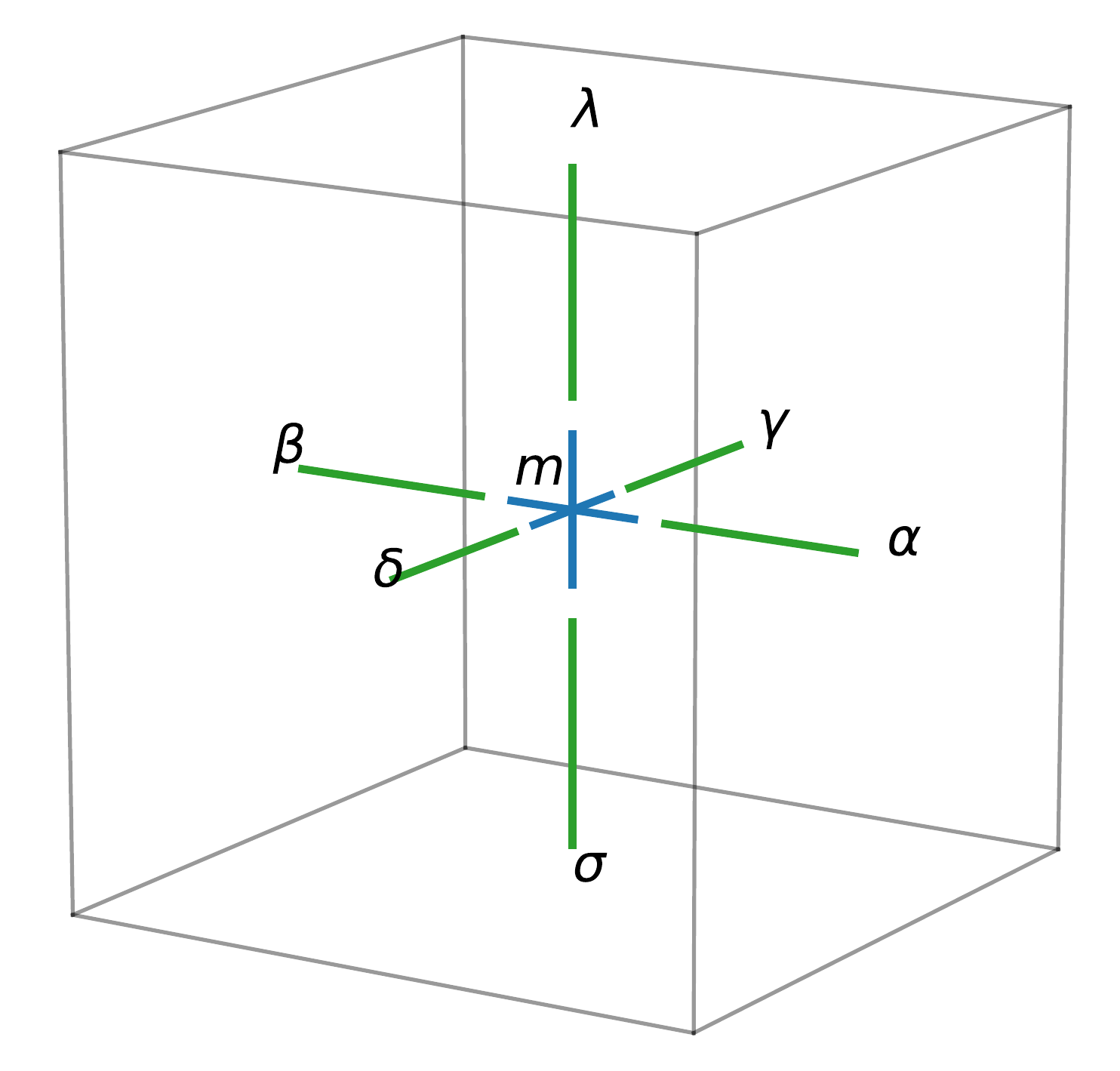}
    \caption{An illustration of the fundamental local tensor as defined in Eq.~\eqref{eq:tensor}.  Here it is drawn inside of a basic cell of the original lattice, and a blue cross at the center shows the dual site associated with this cell.}
    \label{fig:basic_tensor}
\end{figure}

The dual variables from the previous section can be used straightforwardly to construct a local tensor from which the entire partition function can be reconstructed.  This formulation is not unique, and a tensor formulation for 3D $U(1)$ was put forth in Ref.~\cite{PhysRevD.88.056005}.  Here we present a different formulation with more symmetry. To form a tensor we first notice that the partition function from Eq.~\eqref{eq:dualpf_st} describes a theory of integer fields located on the sites of a lattice with nearest neighbor interactions.  To isolate the integer fields on the sites, we interpret the Bessel function weights as matrices in their $m$ indices, and factorize them as,
\begin{equation}
    I_{m-m'}(\beta) \equiv A_{m m'}(\beta) = \sum_{\alpha=-\infty}^{\infty} L_{m \alpha}(\beta)
    L^{T}_{\alpha m'}(\beta).
\end{equation}
This decomposition is not unique, and is simply the matrix square-root.
This decouples the integer fields at the sites from their nearest neighbor interaction, and replaces it with an intermediate sum over states.  To form a local tensor we define,
\begin{equation}
\label{eq:tensor}
    T_{\alpha \beta \gamma \delta \lambda \sigma} = \sum_{m=-\infty}^{\infty}
    L_{m \alpha} L_{m \beta} L_{m \gamma} L_{m \delta} L_{m \lambda} L_{m \sigma}
\end{equation}
which is a function of both the spatial and temporal gauge couplings.  An illustration of this tensor can be seen in Fig.~\ref{fig:basic_tensor}.
Note that contracting this tensor geometrically in the shape of the cubic lattice reconstructs the partition function exactly, since through each contraction the Bessel function weights are reconstructed.  Therefore one is to think of each Greek index in Eq.~\eqref{eq:tensor} as being associated with one of the six directions of a cubic lattice.

\begin{figure}[t]
    \centering
    \includegraphics[width=8.6cm]{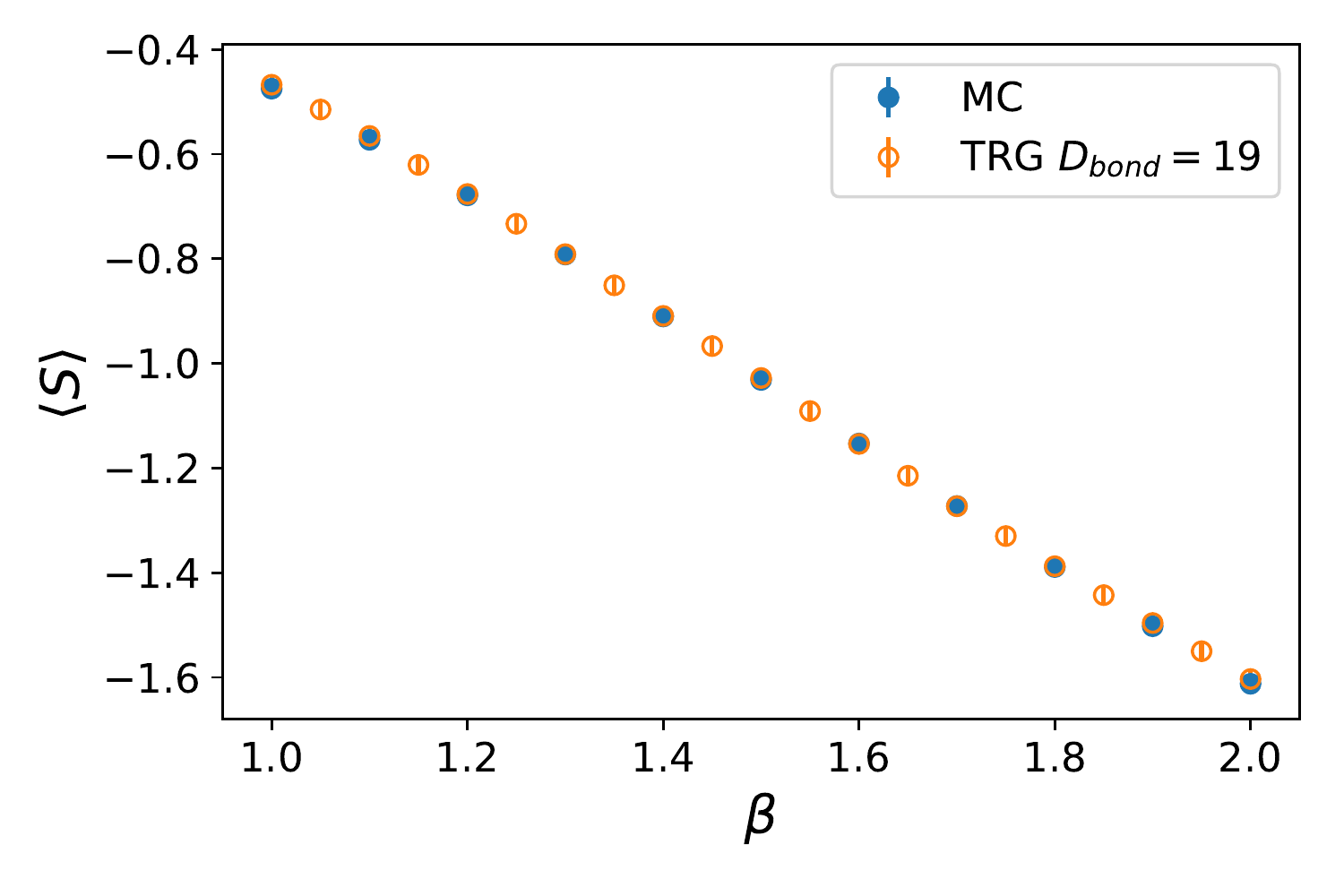}
    \caption{A comparison between Monte Carlo and TRG calculations of the average action per plaquette for a $16^{3}$ lattice.  Here an initial bond dimension, $D_{bond}$, of three was used, and a final $D_{bond}$ of 19.  The error for the TRG was estimated from the largest difference in three different bond dimensions: 15, 17 and 19.  The Monte Carlo calculation averaged over 10,000 configurations and the errors were estimated though jack-knife binning.}
    \label{fig:mc-trg}
\end{figure}

In order to check the validity of the tensor formulation presented here, we compared calculations of the average action per plaquette between Monte Carlo and the TRG,
\begin{equation}
    \langle S \rangle = -\frac{\beta}{3V}\frac{\partial \ln(Z)}{\partial \beta}.
\end{equation}
where $V$ is the spacetime volume.  The Monte Carlo calculations implemented the heat bath algorithm on the weights from Eq.~\eqref{eq:dualpf_st}.  These calculations were compared with Monte Carlo calculations done in the original field variables from Eq.~\eqref{eq:orig-pf}.  A comparison between Monte Carlo calculations and the TRG can be seen in Fig.~\ref{fig:mc-trg}.
The TRG data was extracted from the numerical derivative with respect to $\beta$ of $\ln(Z)$, which is straight forward to calculate using the TRG.  The error bars on the TRG data were calculated using three different final bond dimensions: 15, 17, and 19; however, these calculations were done by restricting the bond dimension to three states in the initial tensor.  We used the largest difference in the average action between the three data sets to estimate the error and assumed that this largest difference was a good approximation for the error for all points, with the addition of the error from the numerical derivative.  Overall we find good agreement between the two methods, which lends support to the validity of the tensor formulation and calculations.

\section{Continuous-time limit}
\label{sec:contime}
\begin{figure}
    \centering
    \includegraphics[width=8.6cm]{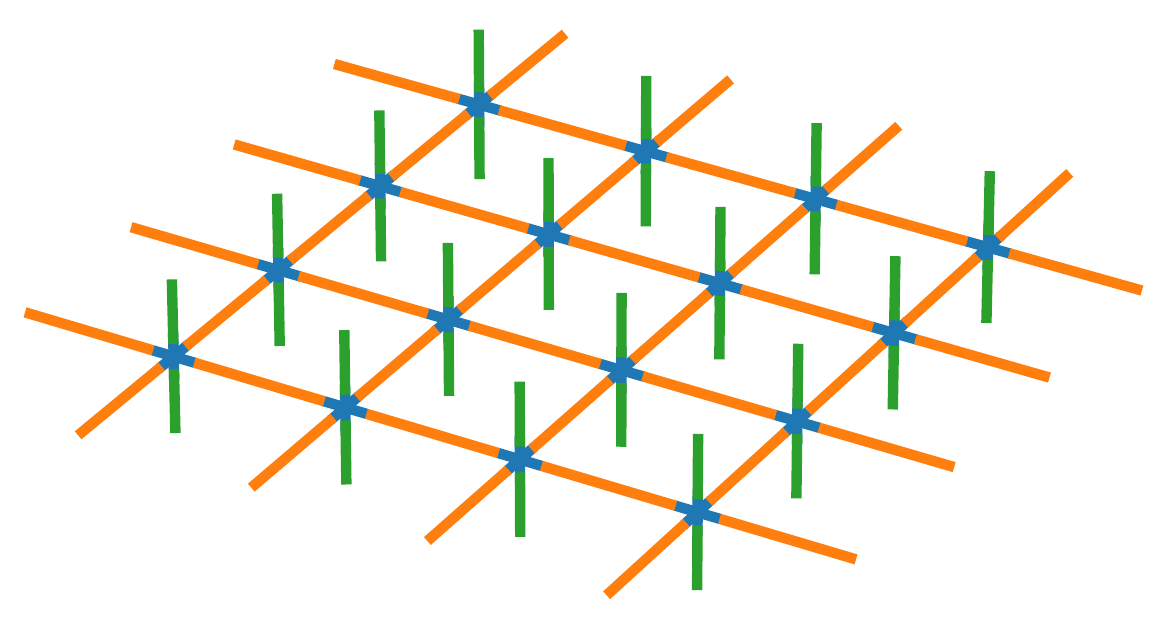}
    \caption{An illustration of the transfer matrix as constructed by local tensors.  Each complete $A$ matrix is orange, while each open index is green.  The remaining open indices are either pointing forward, or backwards in time.  Here it is assumed the lattice continues in both spatial directions.}
    \label{fig:tm}
\end{figure}

Using the tensor formulation of the model, we can construct a transfer matrix.  This is accomplished by contracting local tensors together along a time-slice.  Using periodic boundary conditions, this leaves only tensor indices in the positive and negative time directions.  This construction can be seen in Fig.~\ref{fig:tm}.
In the figure, the tensor contractions have been drawn to represent the ideal case; however, in practice one must truncate and approximate the local basis using some approximation scheme.  Here we used the HOTRG \cite{PhysRevB.86.045139}.  To identify the matrix elements of the transfer matrix we rewrite the action of the model in a slightly different way than before as \cite{RevModPhys.51.659}
\begin{equation}
    S = -\sum_{t} L(t)
\end{equation}
with
\begin{align}
\nonumber
    L(t) = & \frac{1}{2} \sum_{\langle i j \rangle} \ln I_{m_i(t) - m_j(t)}(\beta_{\tau}) \; + \\ \nonumber
    &\frac{1}{2} \sum_{\langle i j \rangle} \ln I_{m_i(t+1) - m_j(t+1)}(\beta_{\tau}) \; +
    \\ &\sum_{x} \ln I_{m_x(t) - m_x(t+1)}(\beta_s),
\end{align}
and now
\begin{equation}
    Z = \sum_{\{m\}} e^{-S}.
\end{equation}
Then the partition function is written essentially as a product of matrices, each of which is associated with a time-slice, and whose indices are the $m$ variables.

Fist consider the diagonal entries of the transfer matrix.  In that case we find
\begin{equation}
L(t) = \sum_{\langle ij \rangle} \ln I_{m_i - m_j}(\beta_{\tau}).
\end{equation}
Next, consider a single change between two time-slices of either $\pm 1$,
\begin{align}
\nonumber
    L(t) = & \frac{1}{2} \sum_{\langle i j \rangle} \ln I_{m_i(t) - m_j(t)}(\beta_{\tau}) \; + \\
    &\frac{1}{2} \sum_{\langle i j \rangle} \ln I_{m_i(t+1) - m_j(t+1)}(\beta_{\tau}) +
     \ln I_{1}(\beta_s).
\end{align}
This is the first off-diagonal contribution.  One then proceeds systematically through all possible changes in the $m$s to identify the matrix elements.

In order to relate this model to a quantum Hamiltonian in two spatial dimensions we must find a limit for this transfer matrix where,
\begin{equation}
\label{eq:tm-expand}
    \mathbb{T} \simeq \mathbf{1} - a H + \ldots
\end{equation}
with $a$ the temporal lattice spacing, and $H$ a Hamiltonian.
Here we ignore an overall thermodynamic constant and work with normalized Bessel functions, as used in Refs.~\cite{jin:2018,judah:2018}.  We normalize the Bessel functions by the zeroth order Bessel function, $t_n(z) \equiv I_n (z) / I_0 (z)$.  These have the following behavior for large and small arguments,
\begin{align}
    t_n(z) &\simeq 1 - \frac{n^{2}}{2 z} + \mathcal{O}(z^{-2}) \quad \text{for $z \rightarrow \infty$} \\
    t_n (z) &\simeq \frac{z^n}{2} + \mathcal{O}(z^{n+2}) \quad \text{for $z \rightarrow 0$.}
\end{align}

To take the continuous-time limit we imagine forcing the temporal couplings to be very strong so to force uniformity in the time direction and simultaneously we make the temporal lattice spacing very small to approach continuity. To that end we take $\beta_{\tau} \rightarrow \infty$, and $\beta_{s} \rightarrow 0$, and the temporal lattice spacing, $a \rightarrow 0$ such that
\begin{equation}
\label{eq:consts}
	U \equiv \frac{1}{\beta_{\tau} a}, \quad
    X \equiv \frac{\beta_{s}}{a}
\end{equation}
are kept constant, and we keep terms in the expansion of the normalized Bessel functions that are of $\mathcal{O}(\beta_s)$ and $\mathcal{O}(\beta_{\tau}^{-1})$.
This gives a transfer matrix that implies a Hamiltonian of the form
\begin{equation}
\label{eq:ham}
	H = \frac{U}{2} \sum_{\langle i j \rangle} (L_{i}^{z} - L_{j}^{z})^{2} - X \sum_{i} U^{x}_{i},
\end{equation}
with the sum $\langle i j \rangle$ over nearest-neighbor pairs.  The operators in this Hamiltonian are defined as follows in the $z$-component of angular momentum basis, as they are in Refs.~\cite{jin:2018,judah:2018},
\begin{align}
    &L^{z}|m\rangle =  m|m\rangle \\
    &U^{x} = \frac{1}{2}(U^{+} + U^{-}) \\
    &U^{\pm}|m\rangle = |m \pm 1 \rangle.
\end{align}
These operators satisfy the commutation relations $[L^{z}, U^{\pm}] = \pm U^{\pm}$, $[U^{+},U^{-}] = 0$.  We see the first term favors ``aligning'' adjacent rotors, while the second term attempts to disorder and scramble the rotors.  


\subsection*{Calculations of the ground state energy}

For small systems it is possible to calculate the energy eigenvalues of Hamiltonian \eqref{eq:ham} accurately using the TRG, and compare with perturbation theory calculations.  Re-scaling, and using
\begin{align}
    H_{0} &= \frac{1}{2}\sum_{\langle i j \rangle} (L_{i}^{z} - L_{j}^{z})^{2} \\
    V &= -x \sum_{i} U^{x},
\end{align}
with $x = X/U = \beta_{s} \beta_{\tau}$, we see the ground state for the un-perturbed Hamiltonian is infinitely degenerate.  We add a term to break this degeneracy, and then remove this contribution at the end if the answer permits.  Then,
\begin{equation}
    H_{0} = \frac{1}{2}\sum_{\langle i j \rangle} (L_{i}^{z} - L_{j}^{z})^{2} + h\sum_{i} (L_{i}^{z})^{2},
\end{equation}
which picks out the $m=0$ state as the ground state for $H_{0}$.  Note that in the case of spatial open boundary conditions, this state is picked out automatically.   We can use the perturbative formulae for the $n$\textsuperscript{th} energy eigenvalue \cite{RevModPhys.51.659},
\begin{equation}
    E_{n} = \varepsilon_{0} + x \varepsilon_{1} + x^{2} \varepsilon_{2} +\ldots
\end{equation}
with
\begin{align}
    x\varepsilon_{1} &= \langle n | V | n \rangle \\
    x^2\varepsilon_{2} &= \langle n |V g V |n \rangle \\ 
    x^3 \varepsilon_{3} &= \langle n | V g V g V |n \rangle - \langle n | V | n \rangle \langle n |V g^2 V |n \rangle \\ \nonumber
    x^4 \varepsilon_{4} &= \langle n | V g V g V g V |n \rangle - \langle n |V g V |n \rangle \langle n |V g^2 V |n \rangle \\ \nonumber
    &+ \langle n | V | n \rangle \langle n | V | n \rangle \langle n |V g^3 V |n \rangle \\
    &- \langle n | V | n \rangle \langle n | V g V g^2 V + V g^2 V g V | n \rangle \\ \nonumber
    &\vdots 
\end{align}
and $g = (1 - |n \rangle \langle n |)/(\varepsilon_{0} - H_{0})$ to compute the different energy states.

Consider the perturbative corrections for the ground state energy, \emph{i.e.} $n=0$.  We will restrict the local Hilbert space to three states, a ``spin-1'' system, with $m = \pm 1, 0$ possible at each site.
Noticing that the perturbation $V$ raises or lowers the angular momentum by one, the first contribution must be at second order.  We find,
\begin{equation}
    \varepsilon_{2} = -\frac{1}{4}N_{x}N_{y}.
\end{equation}
Similarly, the next contribution must be at quartic order,
\begin{align}
    \varepsilon_{4} = -\frac{N_{x}N_{y}}{16} \left[ \frac{(N_{x}N_{y}-5)}{2} + \frac{32}{15} \right]
    + \frac{1}{32}N^{2}_{x}N^{2}_{y}. 
\end{align}
The unperturbed ground state energy, $\varepsilon_{0}$ is simply zero.

Using the TRG to compare, we can explicitly take the limit described in the previous section in the local tensor, and perform contractions to build a transfer matrix.  We then extrapolate the results to the continuous-time limit.  To match with perturbation theory, the initial tensor is restricted to three states, however the final bond dimension varied depending on the spatial volume.  In the continuous-time limit, we expect that if we find the eigenvalues of the transfer matrix, $\lambda_{n}$, they are related to the energy eigenvalues through,
\begin{equation}
    E_{n} = -\beta_{\tau} \ln(\lambda_{n}).
\end{equation}
This is because Eq.~\eqref{eq:consts} dictates that the temporal lattice spacing is inversely proportional to $\beta_{\tau}$, and if one works in units of $U$, $a = 1/\beta_{\tau}$.

\begin{figure}[t]
    \centering
    \includegraphics[width=8.6cm]{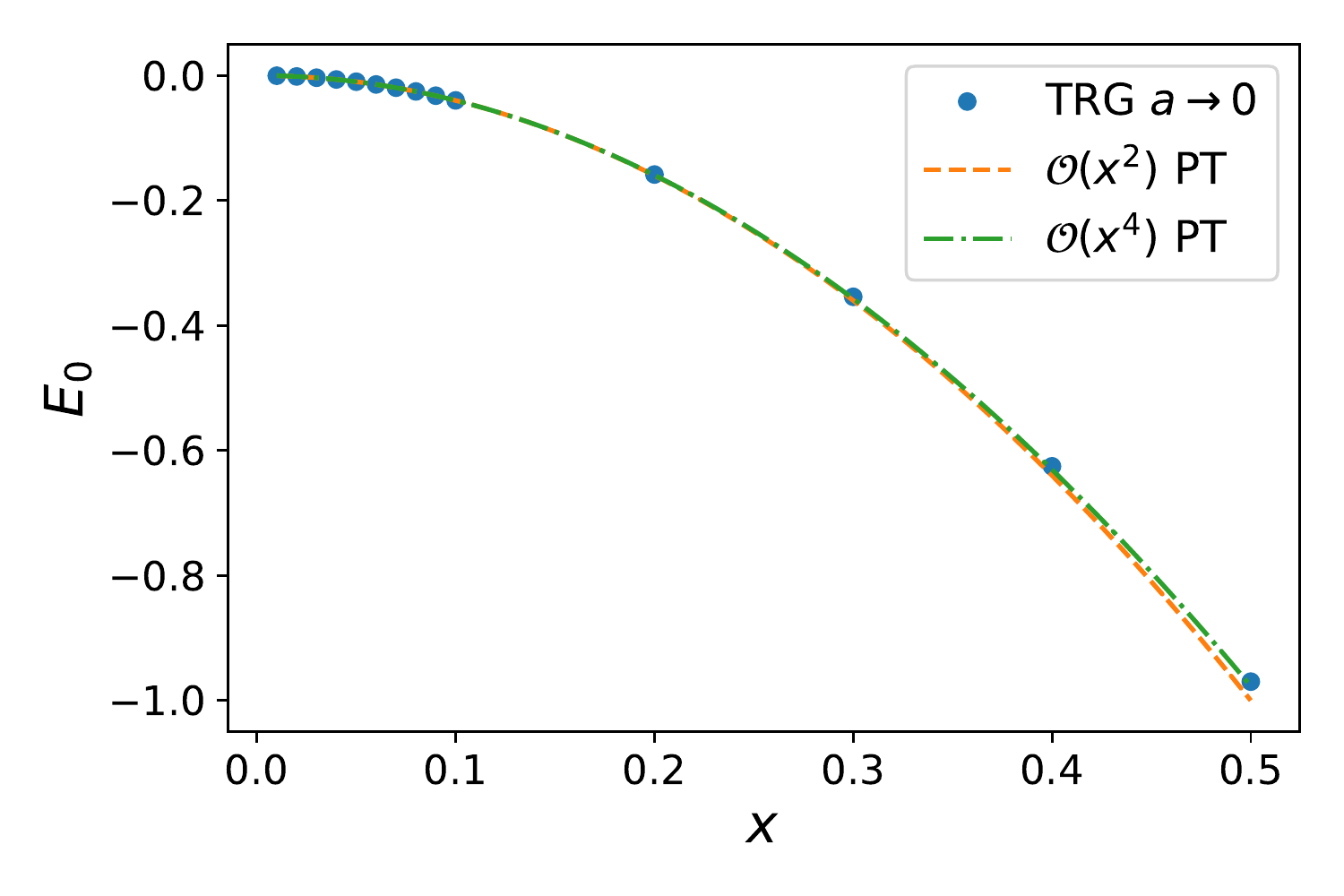}
    \caption{The ground state energy computed using the TRG in the continuous-time limit compared with a perturbation theory calculation of the same quantity to order $x^2$, and $x^4$.  This is on a $4 \times 4$ spatial lattice.}
    \label{fig:gse-dpt}
\end{figure}

A comparison between calculations of the ground state energy using TRG, and using perturbation theory can be seen in Fig.~\ref{fig:gse-dpt}.
Here the two leading-order contributions are plotted, along with data obtained from the TRG calculations extrapolated to the $a \rightarrow 0$ limit.  We find good agreement between the analytic calculation and the numerical calculation with the TRG, indicating that the quantum Hamiltonian does in fact correctly model the $U(1)$ gauge theory with which we started.

\section{Conclusion}
\label{sec:conc}
We have presented a tensor formulation for compact 3D free electrodynamics based on a dual variables formulation for the model.  In order to check this formulation we compared with Monte Carlo calculations done in the dual variables, and the original variables, and found good agreement between the methods.  We used this tensor formulation to extract a quantum Hamiltonian in the continuous-time limit.  In this formulation gauge-invariance in maintained through-out, and the discrete integer fields from the duality transformation can be interpreted as angular momentum quantum numbers in the continuous-time limit, giving a rotor Hamiltonian description for the model.  To check this description, we calculated the ground state energy using the TRG and compared it with a perturbative calculation done with the Hamiltonian and found good agreement.

The Hamiltonian formulation here could be amenable to quantum simulation.  Since gauge invariance is maintained through-out identically, there would be no need to enforce Gauss's law by hand in experiment.  In addition, optical lattice set-ups tailored for Hamiltonians in this basis have already been put forward \cite{jin:2018} and modifications could be straight forward.  We are currently investigating the promise of this approach.

\begin{acknowledgments}
I would like to thank Yannick Meurice and David Kaplan for stimulating discussions, as well as Jack Laiho and Simon Catterall for discussions and for thoroughly reading the paper.  I would also like to thank the organizers at FermiLab for the workshop ``Next steps in quantum science for HEP'' where some of this work was done.  This work is supported by the U.S. Department of Energy, Office of Science, Office of High Energy
Physics, under Award Number DE-SC0009998.
\end{acknowledgments}

\end{document}